# A quantum model for voltage noise: Theory and Experiments

J.I. Izpura

Group of Microsystems and Electronic Materials (GMME)
Universidad Politécnica de Madrid (UPM). 28040-Madrid. Spain.
e-mail: joseignacio.izpura@upm.es

**Abstract**

Although the Fluctuation-Dissipation framework is a first step to get a quantum model for electrical noise, the merging of displacement and conduction currents into the sole current of a series notion like the resistance $R(f)$ does not help in it this task. Used to handle these currents as orthogonal components in electromagnetic waves, the usage of the impedance $Z(jf)=R(f)+jX(f)$ for the noisy device hides the way it interacts with its thermal bath. In contrast to this, the admittance $Y(jf)$ is a parallel notion directly linked with the aforementioned interaction that has led us to develop a discrete model for current fluctuations where single electrons randomly shuttling between any pair of terminals generate the voltage noise we can measure between them.

# 1. Introduction

In 1928, the pioneering scientists J. B. Johnson and H. Nyquist showed that the voltage noise measured between the terminals of any resistor had a thermal origin. The excellent noise measurements of Johnson [1] elegantly explained by Nyquist [2] paved the way to calculate and control the electrical noise of devices and circuits. Concerning noise calculations let us cite a work of J. R. Pierce in 1948 [3] showing that electrical noise in circuits comes from capacitors and inductors rather than from resistors or from their resistances, as it could suggest the noise densities appearing in the circuit of Fig. 1. We mention a single circuit because the two circuits shown in this figure are equivalent circuits that use noise densities having different units.

"Equivalent circuits" means that the electrical magnitudes measured between the two terminals A and B of both circuits will give exactly the same values. Thus, they are equivalent for any external circuit like the front-end electronics of a spectrum analyser connected between terminals A and B. Note that their equivalence is obtained at the cost of having different internal structures, in such a way that their components use to have different currents for example. Readers agreeing with something like: "*from the noise viewpoint these circuits are equivalent only up to a certain degree due to the different dissipation of their resistances*" [4] should keep in mind the above definition and wait for the answer this paper gives to this misconception.

Concerning noise viewpoints a very interesting one is the quantum approach for noise processes that led to the Fluctuation-Dissipation framework published by H. B. Callen and T. A. Welton in 1951 [5]. This is a nice paper we roughly met ten years ago from interactions with colleagues contending that the Fluctuation-Dissipation Theorem derived from [5] was enough to justify any electrical noise in resistors. Used to handle electrical and related noises in low-noise circuits, resistors, capacitors and resonators we enjoyed learning from [5] to conclude that, inspired by the role that the resistance $R(\omega)$ played in the works of Johnson and Nyquist, Callen and Welton showed a quantum framework for noisy processes where any two-terminal device (2TD) with frequency-dependent resistance $R(\omega)$ should exhibit voltage noise in TE. Let us note the use done in [5] of the redundant word "generalized" to distinguish $R(\omega)$, the *frequency-dependent resistance*, from the *flat or constant resistance R* studied at undergraduate levels. As it should be known, resistors bearing a constant $R$ between terminals only is a theoretical notion in our physical world where the lowest dielectric permittivity that can be found between two close terminals like those of a resistor is that of vacuum $\varepsilon_0$ [6].

Thus, no resistors of pure resistance $R$ between terminals exist, as Nyquist likely knew given the ending loads he used to match a lossless transmission line in [2]. Hence, any resistor we can make will be shunted by a capacity $C$ as Johnson already considered in [1]. Since any capacitor of capacity $C$ that we can make will show a differential conductance $G \neq 0$ at temperature T>0 [7], we should consider resistors and capacitors as the same type of 2TD represented by the complex admittance of Fig. 2. Thus, resistors and capacitors offering generalized resistance $R(\omega)$ thanks to their $C$ between terminals should exhibit a voltage noise between them [5], as it is found in experiments. But let us excel this notion by showing that resistors and capacitors are noisy 2TD because they show a complex admittance $Y(j\omega)=G+jB(\omega)$ between their terminals that allow them to interact with the electromagnetic radiation of its thermal bath [8, 9].

Although the work of Callen and Welton appeared in 1951 we have not found other models than [8, 9] giving the mean value of the fluctuations of electrical energy that should give rise to Johnson noise following [5]. To the best of our knowledge, the first model giving this value and conceptually agreeing with the *Irreversibility* word in the title of [5] was shown in [8], whose (Action/Reaction) dynamics agrees well with the (Fluctuation/Dissipation) one of [5]. Added to this, our model is a discrete one that keeps the charge discreteness observed in the noise of nanodevices like those of [10]. This makes it useful to explain noise at the nanoscale, but also other noises due to such granularity like the phase noise or "line-broadening" of the output signal of electronic oscillators [11, 12, 13]. From the familiar displacement and conduction currents, our model explains the way thermal energy generates voltage noise between two terminals at distance $d$ in space regardless of the material or medium between them. To see why this is so, let us begin with a few notions we should take in mind about electrical noise.

**2. Electrical noise in resistors: past, present and revisiting the past**

Fig. 1 shows the series and parallel equivalent circuits for the Johnson noise of a resistor of resistance $R$ in TE at temperature $T$. This is a voltage noise between terminals of spectral density $S_V=4kTR$ V$^2$/Hz that is directly measured by a voltmeter, which in its general meaning is an instrument measuring the instantaneous difference of electrical potentials between two conductors at distance $d$ in space. Regardless of its ability to give spectral densities like $S_V$, a voltmeter measures the aforementioned difference that we call the instantaneous voltage $v(t)$ between two conductors called terminals. Thus, voltmeters cannot directly measure the current noise of Fig. 1-b because they must wait for its conversion into a voltage like $v_n(t)$ and the null waiting time that Fig. 1-b suggests for this conversion is not possible in our physical world (see below).

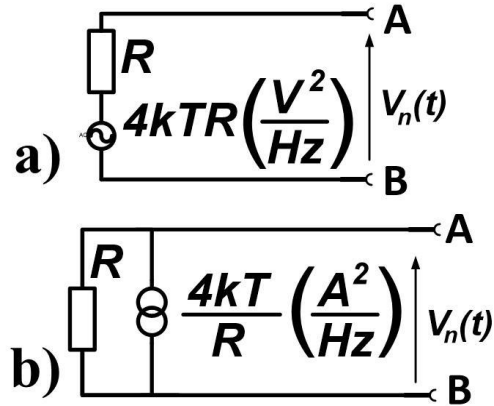

**Figure 1.** Dissipation-based equivalent circuits for electrical noise in resistors: **a)** Series or Thèvenin circuit and **b)** Parallel or Norton circuit.

Resistors are 2TD whose voltage noise in TE is called its Johnson noise and capacitors are similar 2TD whose voltage noise uses to be handled by its mean square value $\langle v^2(t) \rangle$ known as "k$T/C$ noise". For any capacitor of capacity $C$ kept in TE at temperature $T$ this value is: $\langle v^2(t) \rangle = kT/C$ V$^2$ where k is the Boltzmann constant. This k$T/C$ noise of capacitors already was known in 1948 [3], but perhaps less known is the fact that any capacitor at T>0 will show a differential resistance $r_d$ between plates [7]. This means that every capacitor will show a Johnson noise of density $S_V(f)=4kTr_d$ V$^2$/Hz at low-frequencies and whose mean square value will be its k$T/C$ noise.

As shown in the Introduction, the equivalent circuit of resistors and capacitors is the admittance shown below the relaxation cell of Fig. 2. The *relaxation cell* term refers to a space volume $V_{cell}$ between two parallel-plate terminals at distance $d$ where electric fields orthogonal to these plates can exist and relax with time $t$, usually as $exp\{-t/\tau\}$ to simplify. The utility of this cell whose volume can contain solid matter or vacuum as well [7] is shown below. Thus, Fig. 2 is the 1-D model of a homogeneous relaxation cell whose terminals of area: $A_T = w \times h$ at distance $d$ define its volume $V_{cell} = w \times h \times d$ cm$^3$ for $w$, $h$ and $d$ given in cm. For vacuum "filling" this cell a constant electric field $E(t)=1$ V/cm along $d$ would correspond to a constant voltage $v(t)=d$ volts between its terminals. Note that a surface-mount resistor (or capacitor) of length $d$ and cross section $A_T = w \times h$ is a relaxation cell like that of Fig. 2, which in turn is the Cartesian version of other relaxation cells like the cylindrical one of an old carbon resistor.

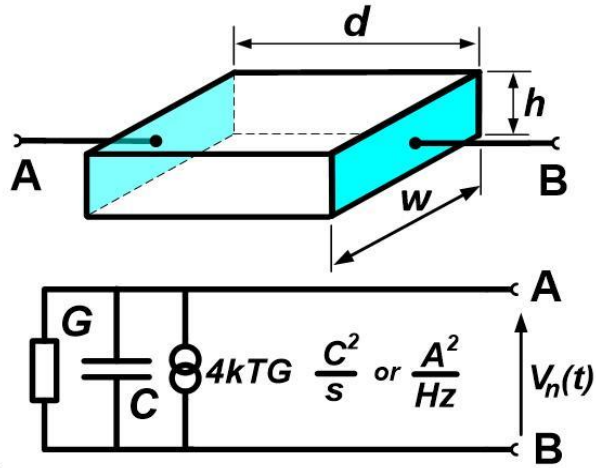

**Figure 2.** Complete model and circuit for noise in resistors and capacitors that uses the charge noise power 4k$TG$ (Coulombs)$^2$/second shown in [8, 9] or the familiar Nyquist density 4k$TG$ A$^2$/Hz deduced from [1, 2]. Note that C$^2$/s and A$^2$/Hz are dimensionally equivalent units.

Thus, every 2TD in general and every resistor in particular is a relaxation cell of some volume $V_{cell} \neq 0$ regardless of its shape, where electric fields can exist and fluctuate with time. Hence, each resistor has a degree of freedom for electrical energy linked with its voltage $v(t)$ and fluctuations of this energy will give rise to a fluctuating voltage $v(t)$ between its terminals as time passes. Before showing that Johnson noise is born in this way, let us fix an old misconception in electrical noise concerning "resistance".

## 3. Demystifying noisy and noiseless resistances

Having read years ago about "ohmic" or noisy resistances and "noiseless" ones like the dynamical resistances of junction diodes, we took this possibility with some care. Knowing that both types of resistances were equivalent to set the bandwidth of an electronic filter, we concluded that electrical noise should not be able to distinguish them. As the real part of $Z(j\omega)$, resistance only is the immaterial ratio $R=V/I$ between amplitudes of sinusoidal voltage (V) and sinusoidal current (I) of the same frequency $f=\omega/(2\pi)$ found in phase in a device. Thus, there is only one type of resistance $R(\omega)$ and the "noisy and noiseless resistances" is a wrong notion that appears when the immaterial magnitude (resistance) is confused with the physical device (resistor), which is the quantum system that shows the voltage noise we measure between its terminals.

Knowing that the immaterial $R$ is not the physical device whose noise we are dealing with it should be clear that the $R$ of a resistor cannot generate its electrical noise as the linear dependence on $R$ of the noise density $S_V$=4k$TR$ V$^2$/Hz of Fig. 1-a could suggest. For similar reasons the ratio $G$=I/V giving the conductance of a resistor also is unable to generate electrical noise as the linear dependence on $G$ of the Nyquist density $S_I$=4kT$G$ A$^2$/Hz of Fig. 1-b might suggest. Thus, the linear dependence of $S_V$ on $R$ and of $S_I$ on $G$ has a hidden meaning that is unveiled when the capacity $C$ of each resistor is considered. This meaning is that "*the mean square voltage noise between terminals of the resistor in TE will be kT/C V$^2$ regardless of the capacity C existing between them*" [8]. Thus, applying thermal equipartition to the capacity $C$ of any resistor in TE we should obtain its Johnson noise, which would be a voltage noise born in its capacity $C$. Next Section shows the relevance of this feature of Johnson noise known years ago [3], but that is not very familiar today.

## 4. Applying thermal equipartition to resistors

Since the degree of freedom having to do with the voltage between terminals of a resistor involves its capacity $C$, let us consider the energy $U_E(v)$ "stored in its $C$" when the voltage between terminals is $v(t)$. This instantaneous energy $U_E(t)$ is:

$$U_E(v) = \frac{Cv^2}{2} \Rightarrow U_E(t) = \frac{C}{2}[v(t)]^2 \qquad (1)$$

Hence, the mean fluctuation k$T$/2 of this energy due to electric fields thermally fluctuating "in $C$" (properly said: "in the volume $V_{cell}$ of the device") would be observed as a randomly varying voltage $v(t)$ whose mean square value <$[v(t)]^2$> should be:

$$\langle U_E(t) \rangle = \frac{kT}{2} = \frac{C}{2}\langle [v(t)]^2 \rangle \Rightarrow \langle [v(t)]^2 \rangle = \frac{kT}{C} \qquad (2)$$

Following [8] the random displacement currents (thus current noise or current fluctuations) driving the admittance of Fig. 2 is what produces the Johnson noise found in resistors. Taking this current noise as impulsive noise to keep the integrity of each displaced electron, its spectral density will be flat: $S_I(f)$=X A$^2$/Hz. Details are given in [9] on the way this integrity is kept by displacements of electrons between terminals where each electron ending this passage appears as a sudden displacement current in $C$. The null transit time ($\tau_T$=0) electrically found when one of such passages ends prevents any "chopping of the charge $q$=1.6×10$^{-19}$ C as it passes", but by no mean it means that its corresponding fluctuation of energy did not took a previous time to take place [9]. This occurs because although the passage of each electron takes a non-null time, no voltage change occurs during this time and just when this passage ends, the voltmeter "feels" an instantaneous voltage step between terminals of $q/C$ volts. This shows why electronics is useless to prevent this impulsive current noise giving rise to the phase noise "close to the carrier" (line broadening) of electronic oscillators: when it feels the voltage step (effect) to react, the fluctuation (cause) already is gone [12].

The mean square voltage noise in the circuit of Fig. 2 collects the contributions at each $f$ of the flat density $S_I(f)$ driving its admittance $Y(jf)$. Since these contributions to the voltage noise is the product of $S_I(f)$ by $|Z(jf)|^2$, the square magnitude of $Z(jf)=1/Y(jf)$, the mean square voltage noise we are looking for is:

$$\langle [v_n(t)]^2 \rangle = X \times \int_{f=0}^{f=\infty} \frac{R^2}{1+\left(\frac{f}{f_C}\right)^2} df = X \times \frac{R^2}{2\pi RC} \times \frac{\pi}{2} = X \frac{R}{4C} \qquad (3)$$

From Eq. (2) and (3) we get: $X=S_I(f)=4kTG$ A$^2$/Hz that is the Nyquist density of Fig. 1-b. This easy result for the electrical noise of resistors and capacitors comes from the truly impulsive current fluctuations we have used to keep the integrity of each displaced electron. Next Section shows the feasibility of this noise from the mean energy required by its fluctuations of energy.

## 5. Monoelectronic fluctuations of energy in resistors

To keep short the length of this paper let us use familiar notions in statistical physics about ionization of donor atoms in n-type semiconductors. We mean their high ionization factor leading to the full-ionization approach when the donor energy level is a shallow one, thus close to the bottom of the Conduction Band (CB). At room $T$ where thermal energy is k$T$≈26 meV, a donor level lying 3 meV below the bottom of the CB allows speaking about a shallow donor level. In this case and for moderate doping levels the donor atoms would show a high ionization factor F($T$). For F($T$)=80% each donor would be ionized the 80% of time on average due to the small energy (U$_d$=3meV≈k$T$/9 at room $T$) that its outer electron would need to become a carrier in the CB.

Going to the dynamics of this process, each ionization becomes a likely event because it is activated by a small energy U$_d$<<k$T$. From the complementary process by which a delocalized carrier loosing energy leaves the CB to be located around a donor ion, we can envisage a detailed balance of these ionization-capture processes sustaining in time the average ionization factor F(T). Calculating the rate of ionization events per unit time in each cm$^3$ of n-type material in our device we would obtain a high value. This high rate agrees with the high likelihood of these events needing U$_d$≈k$T$/9 to take place in our device. With these notions on thermally driven events let us face the likelihood of other electronic events activated not by one tenth of k$T$, but by a *few parts per million* of k$T$.

We refer to the likelihood of the passage of an electron between the terminals of resistors behaving accordingly to the relaxation cell of Fig. 2. To have some figures let us take as an example a resistor of $R$=1kΩ made of lightly doped silicon, with $w \times h$=0.1 cm$^2$ and $d$=1 cm, thus with $V_{cell}$=0.1 cm$^3$. From these values and from the dielectric permittivity $\varepsilon = \varepsilon_r \varepsilon_0$ for silicon, the capacity $C_\varepsilon$ of this resistor would be: $C_\varepsilon \approx$0.1 pF. This capacity is solely due to the dielectric nature of its inner material [6], but when this resistor is put in the plug-in of a noise measuring setup its relaxation cell is modified by any stray capacity $C_{stray}$ due to wiring or to the plug-in itself. Adding $C_{stray}$ to $C_\varepsilon$ we get the total capacity $C=C_\varepsilon+C_{stray}$ for the relaxation dynamics of fluctuations of electric field in our resistor. For $C_{stray} \approx$0.4 pF (typically) we have: $C$=0.5pF.

Thus, our 1kΩ resistor under test would be a relaxation cell whose $\tau$=0.5 ns time constant could be lowered by reducing $C_{stray}$. For $C_{stray}$=0 the lowest value thus obtained for $\tau$ is $\tau_d$, the Maxwell relaxation time of the material within the cell: $\tau_d=RC_\varepsilon$ [6]. This is the "device form" of $\tau_d=\varepsilon/\sigma$ linking the dielectric permittivity $\varepsilon$ and the conductivity $\sigma$ of the material in a homogeneous cell. It is worth noting that a cell with $\tau$=0 does not exist because an electric field existing within such a cell would relax in zero seconds $\Delta t \rightarrow$0. For this to be true the energy that was in $V_{cell}$ should be moved out of $V_{cell}$ at infinite speed and this disagrees with Special Relativity. Used to electrodynamics we could think of the "aether" with null $\varepsilon$=0 required to have $\tau_d$=0 in $V_{cell}$. For $\varepsilon$=0<$\varepsilon_0$, the speed of the electromagnetic wave in such aether should surpass the speed of the light in vacuum $c=1/\sqrt{(\varepsilon_0\mu_0)}$.

Once the utility of the relaxation cell has been shown, let us study the likelihood of events like the passage of electrons between terminals in our resistor of $R$=1kΩ that is shunted by its $C$=0.5pF. Thinking of electrons as small corpuscles, their passages through its capacity $C$ between terminals separated by $d$=1 cm in our resistor could be seen as fast jumps of electrons between them that is the picture used in [8]. Due to its $\tau$=0.5 ns time constant, the Johnson noise density $S_V$=4k$TR$ V$^2$/Hz of our resistor only would be flat up to the cut-off frequency of its relaxation cell $f_C=1/(2\pi\tau) \approx$318 MHz. This limit for $S_V$ that we hardly would consider in Fig. 1, is set by equipartition stating that the voltage noise of our resistor at room $T$ is limited to 91 μV$_{rms}$ in order to keep a mean fluctuation of k$T$/2 Joules in its electric field linked to $C$, see Eq. (2).

Since Johnson noise has equal probability for positive than for negative values, thus zero mean, the most likely voltage of $C$ in TE is zero. This leads to consider jumps of electrons between terminals of our resistor when its $C$ is discharged. In this case each jump would set a voltage step in our resistor of amplitude: $\Delta V = q/C = 0.32$ µV that would set "in $C$" the fluctuation of $U_{FL}$ Joules given by Eq. (1) for $v(t) = 0.32$ µV. Using the electronic charge $q$ to express this fluctuation of energy Eq. (1) becomes:

$$U_{FL} = \frac{C}{2}\left(\frac{q}{C}\right)^2 = \frac{q^2}{2C} = q \times \frac{q}{2C} \qquad (4)$$

Note that the $q/(2C)$ factor of Eq. (4) is a voltage giving the size of $U_{FL}$ in eV. For $C = 0.5$ pF we have: $2C = 10^{-12}$ F and $U_{FL} = 1.6 \times 10^{-7}$ eV. This amount of energy is so small that $kT \approx 26$ meV at room $T$ is: $1.6 \times 10^5$ times higher. Admitting that ionizations requiring a tenth of kT occur at high rates, the passages of electrons between terminals of a resistor should occur at a huge rate $\lambda$. From symmetry reasons the average number of electrons randomly arriving in each terminal should be $\lambda/2$ and its associated current would be $I_T = q(\lambda/2)$. This current is counter-balanced (on average) by an equal number of electrons randomly leaving each terminal, thus giving rise to a detailed balance of currents in TE with zero mean (null dc current in the resistor) but with non-null mean square (noise in the resistor). Keeping these notions in mind, let us obtain the mean rate $\lambda$ at which these jumps should take place in resistors.

**6. Detailed balance of electrical powers**

Detailed balance is a fruitful notion in Solid-State physics for processes like emission-capture ones leading to the aforementioned ionization factor F(T). In TE, the number of emissions and captures per unit time in a given volume are equal on average, but not at each instant. This special cancellation of two complementary processes allows for the existence of *equilibrium* (average cancellation) but *noise as well*, due to its non-cancellation at each instant of time. If cancellation was perfect (e. g. at each instant) no noise would be observed and, although complementary processes like these could exist, we hardly could give an empirical proof about. Hence, Johnson noise would prove the existence of two electrical, complementary processes involving electrons in resistors. Though speak about "opposed" emission-capture processes is possible, we prefer the word "complementary" to better express the notion of *orthogonal* processes like these that appear in detailed balances associated to TE.

To better show what we mean let us consider the two complementary flows of energy that would keep a resistor in TE. From the mean square voltage of Eq. (2) for its Johnson noise and the meaning of Joule effect: "*energy arriving in the resistor in electrical form is leaving it as heat*" we have to consider a flow of heat leaving the resistor and an equal flow of electrical energy entering it on average. This means a flow of electrical energy losing its electrical form and a flow of other energy (likely photons) that should be losing its electromagnetic form to become electrical energy entering $V_{cell}$. This led us to recall the way we handle orthogonal currents in an electromagnetic wave for energy that keeps its electromagnetic form (displacement currents) and conduction currents for energy that loses its electromagnetic form or that is dissipated as this wave interacts with matter. From this notion it is not difficult to follow the reasoning to come with electrical powers. The average energy dissipated in our resistor per unit time by its own Johnson noise driving its $R$ will be the mean square voltage of Eq. (2) divided by $R$ (Joule effect). This ratio that is the active power in the resistor will be:

$$P_{act} = \frac{\langle [v_n(t)]^2 \rangle}{R} = \frac{kT}{RC} \qquad (5)$$

For our resistor in TE at room $T$ with $\tau=RC=0.5$ ns, we would have: $P_{act}\approx 8$ pW leaving it as heat and demanding an equal average power entering it from its thermal bath or being generated in its $V_{cell}$ by this bath. Let us call $P_{react}$ this electrical power that likely will be born in a reactive element. If the electrical power that a dynamo delivers is generated in its inductors that are placed (or displaced) in space so as to undergo fluctuations of magnetic field, the electrical power generated in resistors should not depart from this notion where fluctuations of one of the electromagnetic fields generate electrical power in a reactive element linked with such field. Since the capacity $C$ of a resistor is linked with the electric field of its relaxation cell, fluctuations of electric field should generate the electrical power we observe as its Johnson noise. Hence, resistors should collect electrical power from their thermal bath by a series of photons that they would absorb in their relaxation cell. These photons would be those low-energy ones absorbed by individual electrons to pass between terminals.

To continue, let us recall that electrical energy can be generated and stored in a volume of space like $V_{cell}$, but not in its immaterial $C$ and once stored in $V_{cell}$ (once it has lost its electromagnetic form) this energy should be dissipated in $V_{cell}$ as well, not in its immaterial resistance $R$. We write this because we were tempted to say: "*resistors sense* *their temperature by the random series of low-energy photons they collect by their capacity C between terminals acting like an antenna loaded by its own resistance R, which dissipates the energy thus collected*". Educated to assign these types of roles to $C$ and $R$ we have to recall ourselves the immaterial nature of a circuit and this helps us to understand comments like [4]. Note the analogy between displacement and conduction currents sharing the electric field in the electromagnetic wave and the displacement and conduction currents sharing the voltage noise $v_n(t)$ in the admittance of Fig. 2.

Going back to $P_{react}$, it should be the reactive power in the circuit of Fig. 2 driven by the aforementioned fluctuations of electric field, thus driven by the "charge noise in $C$" due to electrons randomly passing between terminals. This leads to the charge noise power of $4kT/R$ Coulombs$^2$/second shown in Fig. 2. Going to the mean power brought to $V_{cell}$ by the $\lambda$ fluctuations of electric field reaching it on average each second it should be $\lambda$ times the mean energy $U_{FL}$ of Eq. (4): $P_{react}=\lambda \times U_{FL}$. This sum in power assumes that these fluctuations of electric field in the space occupied by $V_{cell}$ are uncorrelated, as it suggests the strong negative feedback of this absorption of photons due to the rapid growth of the energy needed to displace an electron increasing a voltage already built in $C$ by previous passages. We mean that to pass from A to B terminals if $C$ is discharged ($v_n(t)=0$ in Fig. 2) an electron needs $U_{FL}$, but a next electron doing a similar passage shortly after would need: $3U_{FL}=4q^2/(2C)-q^2/(2C)$.

Thus the probability of a second passage from A to B terminals is lower than the probability of this next passage occurring in the opposed sense (from B to A terminals). This asymmetry rapidly increases as $v_n(t)$ departs from $v_n(t)=0$, although the huge rate $\lambda$ we are looking for allows the noticeable departures giving rise to noise peaks from time to time. Another reason for the full randomness of these fluctuations is that the random voltage they build "in $C$" mimics the way a random velocity is built "in the mass" of a particle undergoing Brownian motion [14]. Moreover, only a full randomness both in time and sign of these fluctuations in different regions of space (that can be very close: think of a cylindrical resistor of resistance $R$ made from a large number n of slices, each of resistance $R/n$) could explain why the series or parallel connection of resistors does not give rise to any coherent generation of voltage between the terminals of the new device thus obtained. Therefore, discarding correlation or admitting the full randomness of these thermal fluctuations of electric field in space (at least from the viewpoint of their long term average, see below), the balance of $P_{react}$ and $P_{act}$ to keep the resistor in TE leads to:

$$P_{reac} = \lambda \frac{q^2}{2C} = P_{act} = \frac{kT}{RC} \Rightarrow \lambda = \frac{2kT}{q^2 R} = \frac{2kTG}{q^2} \qquad (6)$$

Eq. (6) shows that the average rate of fluctuations in a resistor is a truly huge number, proportional to its conductance $G=1/R$ and to its temperature $T$. For a resistor of $R=1k\Omega$ at room $T$, this rate is: $\lambda \approx 3 \times 10^{14}$ fluctuations per second and the long term average we have just mentioned could start in the picosecond range. This huge $\lambda$ agrees well with the minute energy each electron needs to pass by a fluctuation of electric field between terminals of a resistor and the small amplitude $\Delta V=q/C$ of voltage steps taking place at this huge rate $\lambda$ explains why electrical noise looks like continuous voltage for today's oscilloscopes.

**7. The quantum model at work**

From the $\lambda/2$ electrons per second randomly arriving in each terminal the two equal but opposed currents that mutually cancel on average in a resistor in TE will be: $I_T=q(\lambda/2)$. From Eq. (6) and the thermal voltage $V_T=kT/q$ we get: $I_T=V_T/R=GV_T$. These currents that recall the two opposed $I_{sat}$ of a diode, give a new picture on Johnson noise. As the independent currents they are their shot noise density should be: $S_I=2\times(2qI_T)$ $A^2/Hz$ and from $I_T=GV_T$ we obtain: $S_I=4kTG$ $A^2/Hz$ that is the Nyquist noise density of Fig. 1.b. Thus, the voltage noise that we call Johnson noise in resistors comes from their impulsive shot noise in $C$ called Nyquist noise. But due to the presence of $C$ this current noise only is a theoretical notion and not a current noise that could be "extracted" from the resistor by an I-V converter. Its conversion to voltage noise "already has been done by $C$" when such a converter tries to extract it from the resistor [9].

Converting to voltage noise the shot noise density $S_I=2\times(2qI_{sat})$ $A^2/Hz$ of the two $I_{sat}$ of a diode in TE by the square of its dynamical resistance $r_d$, we obtain its voltage noise under open circuit conditions, which is: $S_V=4kTr_d$ $V^2/Hz$. This is nothing but the Johnson noise of $r_d$ that by no mean is noiseless as we stated in Section 2. However, the misconception shown in Section 2 remains and its dramatic confirmation is the inability of the simulator PSPICE to give the right voltage noise of this noise in a junction diode in TE. We mean the voltage noise of a diode driven by a null dc current source ($I_{DC}=0$). Since PSPICE only considers voltage noise in resistances like the series one of the diode (a small one) and takes $r_d$ as "noiseless" [8], it largely underestimates the voltage noise of the diode in TE, being this a result that can be easily checked by a laptop computer.

Let us say that the "*Thermal_Action-Device_Reaction*" dynamics of [8] has been adapted to the "Fluctuation-Dissipation" framework of [5] that from a circuit viewpoint shows the need of a complex immitance to handle noise in 2TD. This important result confirms something that was well-known in circuit theory: that to handle fluctuations of electrical energy in a circuit one needs reactive elements in it and to handle dissipation of this energy resistive elements are required. Thus, handling together fluctuations and dissipations of electrical energy requires using together both types of elements. This can be done by the impedance *Z(jω)* used in [5] for the Johnson noise of resistors or by the admittance *Y(jω)* of Fig. 2 whose advantages over *Z(jω)* have been shown.

Finally, a striking feature of our model is the easy way it explains and predicts the electrical noise of nanodevices like the small storage capacitor (SC) of a DRAM cell of capacity $C_{SC}$=20 aF ($20 \times 10^{-18}$ F) accordingly to the footnote of Fig. 2 of [10], where the good agreement these authors find between the mean square voltage noise of their SC and $kT/C_{SC}$ is not a surprise for our model. Note that the two terminals used to define this capacity are the bit line (BL) and the MOSFET channel that also is used as a highly sensitive probe of "*single electrons shuttling between these two terminals*" [10]. The most striking result of [10] (see its title) is in its Fig. 1-d where we can track in time the random shuttling of electrons between the two terminals of $C_{SC}$. Counting passages of electrons back and forth in this figure we obtain 50 passages in the 10 seconds span of this graph. This is an average rate $\lambda_{exp}$=50/10=5 passages per second (or fluctuations of energy per second) in this $C_{SC}$ that being at *T*=300K will be shunted by a dynamical resistance *R* [7], thus giving rise to a lorentzian noise spectrum like that of our Eq. (3).

As shown in the footnote of Fig. 1-d the voltage $V_{WL}$ applied to the Word Line of the DRAM cell in this case was: $V_{WL}$=-2.3V and going to the Lorentzian spectrum of noise of Fig. 3-a of [10] for *T*=300K and $V_{WL}$=-2.3V its cut-off frequency is $f_C=10^{-1}$ Hz. From $f_C=1/(2\pi R C_{SC})$ the *R* that shunts $C_{SC}$ in these conditions is: $R=10^{18}/(4\pi)$ Ω. Using these values of $C_{SC}$, *R* and *T* in our Eq. (6) we get λ=4 monoelectronic fluctuations of energy per second, thus very close to the measured rate $\lambda_{exp}$=5 s$^{-1}$. To our knowledge, no noise model exists (except ours) being able to give this quick assessment of electrical noise reaching the nanoscale. To compensate for these nice empirical data let us say that the resistance *R* these authors deduce from their data is taken as "*the channel resistance connected to the SC*" [10]. This proposal is wrong and unnecessary because the meaning of *R* is that of our model: the unavoidable conductance *G*=1/*R* shunting the capacity $C_{SC}$ that these authors elegantly control with the voltage $V_{WL}$.

Going back to Fig. 3-a of [10], the Lorentzian spectrum of noise for $T$=300K and $V_{WL}$=-2.5V shown there has a cut-off frequency of $f_C$=10$^{-2}$ Hz. Following our model this allows to predict that the new rate of fluctuations of energy in $C_{SC}$ should be ten times lower because the shunting or leaking resistance $R$ has been increased roughly by ten as the charge stored in the electrode BL of $C_{SC}$ is better isolated from the outside world by increasing changed $V_{WL}$ from $V_{WL}$=-2.3V to $V_{WL}$=-2.4V. This prediction is done by keeping the k$T/C_{SC}$ of this small capacitor, thus from equipartition that is a key feature of our model. We dare to do it because we know that our model works even for smaller capacitances than the 20 aF of this SC. We refer to each differential capacitor isolating a conductive channel like that of a FET transistor. The k$T/C$ noise conservation of these differential capacitors, each shunted by a different resistance and modulating the conductive channel was all we needed to explain the ubiquitous 1/$f$ excess noise of solid-state devices in [6].

**Conclusion**

We have presented the first discrete model for the voltage noise in resistors called Johnson noise. The model is valid when thermal equipartition holds, but it can be the starting point for discrete models where the thermal energy k$T$ no longer surpasses the interaction energy of the noisy device with the electromagnetic radiation of its thermal bath. This model that was born to explain the ubiquitous 1/$f$ resistance noise of solid-state devices, equally works to explain flicker noise of vacuum devices, voltage noise in nanodevices and other noises having to do with the discreteness of the electric charge like the phase noise of electronic oscillators.

**Acknowledgements**

This work was supported by the E. U. project Nº 304814 RAPTADIAG and by the MAT 2013-45957-R project of the Spanish Ministry.